\pdfoutput=1
\documentclass[12pt,a4]{article}
\usepackage{relsize}
\usepackage{array}
\usepackage{a4wide}
\usepackage{latexsym}
\usepackage{cite}
\usepackage{mathrsfs}
\usepackage{amssymb}
\usepackage{amsmath}
\usepackage{graphicx}
\usepackage{enumerate}
\usepackage[usenames,dvipsnames]{xcolor}
\usepackage{todonotes}
\definecolor{MyDarkBlue}{rgb}{0.15,0.15,0.45}
\usepackage[linktocpage=true]{hyperref}
\hypersetup{
colorlinks=true,
citecolor=MyDarkBlue,
linkcolor=MyDarkBlue,
urlcolor=MyDarkBlue,
pdfauthor={},
pdftitle={},
pdfsubject={hep-th}
}
\usepackage{multirow}

\newcommand{\CP}{\mathbb{CP}}

\newcommand{\be}{\begin{equation}\label}
\newcommand{\ee}{\end{equation}}
\newcommand{\bea}{\begin{eqnarray}\label}
\newcommand{\eea}{\end{eqnarray}}

\makeatletter
\newcommand*{\textoverline}[1]{$\overline{\hbox{#1}}\m@th$}
\newcommand*\bigcdot{\mathpalette\bigcdot@{.65}}
\newcommand*\bigcdot@[2]{\mathbin{\vcenter{\hbox{\scalebox{#2}{$\m@th#1\bullet$}}}}}
\makeatother

\numberwithin{equation}{section}       

\newcommand{\ii}{\mathrm{i}}
\newcommand{\dd}{\mathrm{d}}

\date{}
\begin{document}

\renewcommand{\thefootnote}{\fnsymbol{footnote}}

   \vspace{1.8truecm}

 \centerline{\LARGE \bf {\sc  Non-Lorentzian M5-brane Theories }}
  \vskip 12pt
 
 \centerline{\LARGE \bf {\sc  from Holography} }
 
\vskip 2cm

  \centerline{
   {\large {\bf  {\sc N.~Lambert,${}^{\,a}$}}\footnote{E-mail address: \href{neil.lambert@kcl.ac.uk}{\tt neil.lambert@kcl.ac.uk}}     \,{\sc A.~Lipstein$^{\,b}$}\footnote{E-mail address: \href{mailto:arthur.lipstein@durham.ac.uk}{\tt arthur.lipstein@durham.ac.uk}}  {\sc    and P.~Richmond${}^{\,a}$}}\footnote{E-mail address: \href{mailto:paul.richmond@kcl.ac.uk}{\tt paul.richmond@kcl.ac.uk}}  }    
\vspace{1cm}
\centerline{${}^a${\it Department of Mathematics}}
\centerline{{\it King's College London, Strand, WC2R 2LS, UK}} 
  
\vspace{1cm}
\centerline{${}^b${\it Department of Mathematical Sciences}}
\centerline{{\it Durham University, Durham, DH1 3LE, UK}} 

\vspace{1.0truecm}

 
\thispagestyle{empty}

\centerline{\sc Abstract}
\vspace{0.4truecm}
\begin{center}
\begin{minipage}[c]{360pt}{
M-theory on $AdS_7 \times S^4$ admits a description where the $AdS_7$ factor is constructed as a timelike Hopf fibration over a non-compact three dimensional complex projective space $\tilde{\CP}^3$ \cite{Pope:1999xg}. We consider the worldvolume theory for M5-branes  at a fixed $\tilde{\CP}^3$ radius which, after reduction   along the timelike fibre,  is given by an $\Omega$-deformed Yang-Mills theory with eight supercharges.  Taking the radius to infinity then induces a classical RG flow. We construct the fixed point action which has an enhanced  24 supercharges and which can be understood as  the $(2,0)$ theory of M5-branes on flat space reduced along a compact null Killing direction. 

 }

\end{minipage}
\end{center}

\newpage 
 
\renewcommand{\thefootnote}{\arabic{footnote}}
\setcounter{footnote}{0}

\pagebreak
\tableofcontents

\allowdisplaybreaks

\section{Introduction}

Understanding the dynamics of M-theory is very challenging because its stable objects are not strings but rather two and five-dimensional objects known as M2-branes and M5-branes. A very useful tool for addressing this question is the AdS/CFT correspondence, which relates the conformal field theory living on a stack of M2- or M5-branes to M-theory on $AdS_4 \times S^7$ or $AdS_7 \times S^4$, respectively \cite{Maldacena:1997re}. The conformal field theory for M2-branes is known and for an arbitrary number of branes has a lagrangian description in terms of a Chern-Simons-matter theory with 24 manifest superconformal symmetries and superconformal group  $OSp(6|4)$ \cite{Aharony:2008ug}. In his description, there is a ${\mathbb Z}_k$ action, where $k$ is the quantised Chern-Simons level, that allows a weak coupling limit. On the gravity side the ${\mathbb Z}_k$ acts naturally on the spacelike Hopf fibration of the $S^7$ and $k \rightarrow \infty$ leads to IIA string theory on $AdS_4 \times {\CP}^3$ with constant dilaton. 

Much can be learned about the non-abelian M5-brane theory by dimensionally reducing it to five-dimensional super-Yang-Mills theory. There the Kaluza-Klein modes are interpreted as  solitons so that the extra dimension is recovered non-perturbatively and it has been argued that this is a complete description \cite{Douglas:2010iu,Lambert:2010iw} (timelike and null reductions were considered in \cite{Hull:2014cxa,Lambert:2011gb}). While five-dimensional super-Yang-Mills can be used to compute BPS quantities in the M5-brane theory (see for example \cite{Kim:2011mv,Kallen:2012zn,Kim:2012qf,Minahan:2013jwa,Kim:2013nva}), it remains unclear that it provides a complete description of the M5-brane theory since the five-dimensional theory contains UV divergences \cite{Bern:2012di}, although these could be resolved by   soliton effects \cite{Papageorgakis:2014dma}.  In \cite{Kim:2012tr} an ABJM inspired approach  was adapted to M5-branes  by invoking an ${\mathbb R}\times {\CP}^2$ reduction of ${\mathbb R}\times S^5$, leading to a novel Yang-Mills theory with 12 supersymmetries and a Chern-Simons term. 

More recently, it was shown that rescaling five-dimensional super-Yang-Mills in a non-Lorentzian way induces a classical RG flow whose fixed point is a theory with 24 superconformal symmetries \cite{Lambert:2019evt} and which leads to the DLCQ prescription for the $(2,0)$ theory \cite{Aharony:1997th,Aharony:1997an}. 
In this paper, we  provide a holographic realisation of this phenomenon also inspired by the ABJM construction. Following \cite{Pope:1999xg}, we consider $AdS_7$ as a timelike Hopf fibration over a non-compact complex projective space $\tilde{\CP}^3$ and consider M5-brane embeddings at constant $\tilde{\CP}^3$ radius.  Reducing along the timelike fibre at finite radius then gives a five-dimensional Yang-Mills theory with eight supercharges whose terms are rescaled in a non-Lorentzian way by the radius and modified by an $\Omega$-deformation  \cite{Nekrasov:2002qd}. Taking the radius to infinity then induces a classical RG flow whose fixed point is a theory with 24 superconformal symmetries which can be interpreted as a null reduction of the M5-brane theory in the boundary of $AdS_7$. We conjecture that the boundary theory is UV complete and provides a lagrangian description for the non-abelian M5-brane theory. 

The rest of this paper is organised as follows. In section \ref{holographic} we describe the $AdS_7\times S^4$ background of M-theory and its timelike reduction to $\tilde{\CP}^3 \times S^4$, as first discussed in  \cite{Pope:1999xg}. In section \ref{finite} we construct the non-abelian worldvolume gauge theory for $N$ M5-branes on $\tilde{\CP}^3$ by reducing a single M5-brane on a timelike circle at finite radius and generalizing the result to obtain a non-abelian Yang-Mills  theory with eight supersymmetries along with an $\Omega$-deformation. In section \ref{boundary} we consider the limit where the M5-branes are taken to the boundary. Here the timelike reduction becomes null. We construct the non-abelian worldvolume gauge theory and show that it has eight supersymmetries plus an additional 16 superconformal symmetries. This provides a holographic realization of non-Lorentzian RG flows recently discussed in \cite{Lambert:2019evt}. Section \ref{conclusion} contains our conclusions. We also have some appendices which contain our conventions and various formulae.

While we were working on this project we were made aware of related unpublished work by S.~Kim, S.~Mukhi and A.~Tomasiello \cite{KMT}.

\section{Holographic Setup}\label{holographic}

According to the AdS/CFT correspondence, the worldvolume theory for a stack of M5-branes is dual to M-theory in an $AdS_7 \times S^4$ background. Following in \cite{Pope:1999xg}, we shall consider $AdS_7$ as a timelike $U(1)$ Hopf fibration over a non-compact complex projective space ${{\tilde  {\CP}} }^3$:
\begin{align} \label{ads7}
\dd s^2_{AdS_7} \ = \ -\frac{1}{4}\left(dx^{+}+e^{\phi}\left(dx^{-}-\frac{1}{2}\Omega_{ij}x^{i}dx^{j}\right)\right)^{2}+\dd s^2_{\tilde{\CP}^{3}} \, ,
\end{align}
where $i=1,2,3,4$ (see appendix \ref{conventions} for further conventions) and the metric on ${{\tilde  {\CP}} }^3$ is
\begin{align}
\dd s^2_{{{\tilde  {\CP}} }^3} \ = \ \frac{R_+^2}4 d\phi^2 + \frac14 e^\phi dx^i  dx^i + \frac{e^{2\phi}}{4}\left(dx^--\frac12 \Omega_{ij}x^idx^j\right)^2 \, .
\end{align}
Note that in this construction $x^+$ has period $4\pi R_+$ and $\Omega_{ij}$ is an antisymmetric tensor with nonzero components $\Omega_{13}=\Omega_{24}=R_+^{-1}$ satisfying
${\Omega}_{ij} = -\frac12 \varepsilon_{ijkl}\Omega_{kl}$ and $\Omega_{ik}\Omega_{kj} = -R^{-2}_+\delta_{ij}$.\footnote{We have chosen slightly different coordinates to that of \cite{Pope:1999xg}: $x^3=R_+y_1, x^4= R_+y_2, x^+=R_+\tau, x^- = R_+(\chi - \frac 12(x_1y_1+x_2y_2))$ and our $AdS_7$ has radius $R_+$. }  

In this paper, we will mainly consider brane embeddings at constant $\phi$. Restricting the metric in \eqref{ads7} to finite $\phi$ then gives
\begin{align}
\label{metricconstantphi}
\dd s^2_{\phi} \ = \ \frac{e^\phi}{4}\left[- e^{-\phi}dx^{+}dx^{+} -2 dx^{+}\left(dx^{-}-\frac{1}{2}\Omega_{ij}x^{i}dx^{j}\right)+ dx^i  dx^i\right] \, .
\end{align}
Reducing the M5-brane theory along $x^+$ therefore corresponds to a timelike reduction,  as studied for example in \cite{Hull:2014cxa}. On the other hand, if we take $\phi \rightarrow \infty$ this becomes 
\begin{equation}
\label{metricboundary}
\dd s^2_{\phi\to\infty} \ = \ \frac{e^{\phi}}{4}\left[ -2  dx^+ \left(dx^--\frac12 \Omega_{ij}x^idx^j\right)  +  dx^i  dx^i \right]\ ,
\end{equation}
and so reducing along $x^+$ at the boundary corresponds to a null reduction, as considered in the DLCQ construction \cite{Aharony:1997th,Aharony:1997an}. We will discuss the details of the finite and infinite $\phi$ reductions in the next sections.

Now we wish to argue that M5-branes embedded at finite $\phi$ preserve eight supercharges, but this becomes enhanced to 24 supercharges at the boundary.  In particular, the amount of supersymmetry preserved by the M5-branes corresponds to the number of $AdS_7$ Killing spinors which are eigenspinors of $\gamma_{\phi}$ \cite{Gutowski:1999iu}. The relevant solutions to the $AdS_7$ Killing spinor equation  were given in \cite{Pope:1999xg} and we include their results in appendix \ref{adskse}, adapted to our coordinates. In total there are 32 independent $AdS_7$ Killing spinors.  Firstly there are eight    ($\epsilon^-_{1,8}$ in the notation of \cite{Pope:1999xg}) which depend on $x^+$ but these are not eigenspinors for any value of $\phi$.  Thus they  would not survive our reduction on $x^+$ in any case. Of the remaining 24  one sees that at finite $\phi$ only eight ($\epsilon^-_{3,5}$) are eigenspinors of $\gamma_\phi$. Thus at finite $\phi$ we expect that the M5-brane worldvolume theory admits eight supersymmetries, all of which survive the reduction on $x^+$. At infinite $\phi$ one finds that the other 16 supersymmetries ($\epsilon^-_{2,7}$ and $\epsilon^-_{4,6}$) become eigenspinors of $\gamma_\phi$. These spinors are independent of $x^+$ but have a non-trivial dependence on  $x^i$ and $x^-$. They are naturally thought of as conformal Killing spinors on the M5-brane. In appendix \ref{kseapp} we derive the corresponding 32 conformal Killing spinors for the six-dimensional boundary metric. Hence, after reducing the M5-brane theory along $x^+$ we expect to obtain a theory with eight supercharges at finite $\phi$ which flows to a theory with 24 supercharges at the boundary. We will confirm this holographic prediction below by first constructing the the theory at finite $\phi$ and carefully taking the limit $\phi\rightarrow \infty$. 

Lastly let us comment on the more traditional embedding of an M5-brane in $AdS_7$. As pointed out in \cite{Pope:1999xg} the usual radial coordinate, $\rho$,  in $AdS_7$ is related to $\phi$ through:
\begin{equation}
e^{\phi/2} \ = \ \frac{e^\rho}{\cos (x^+/2R_+)}\ .
\end{equation}
Placing the M5-branes on surfaces of constant $\rho$ leads to the metric
\begin{align}
\dd s^2_{\rho} \ = \ \frac{e^{2\rho}}{4\cos^2 (x^+/2R_+)}\left[-\cos( x^+/R_+)e^{-2\rho}dx^+dx^+  -2 dx^{+}\left(dx^{-}-\frac{1}{2}\Omega_{ij}x^{i}dx^{j}\right)+ dx^i  dx^i\right] \, .
\end{align}
The problem with this embedding is that the metric depends on $x^+$ and so we cannot directly perform a Kaluza-Klein reduction to map the M5-brane dynamics to that of a five-dimensional Yang-Mills gauge theory. On the other hand in the limit  $\rho\to\infty$ we find the boundary metric
\begin{equation}
\label{minkowski}
\dd s^2_{\rho\to\infty} \ =  \ \frac{e^{2\rho}}{4\cos^2 (x^+/2R_+)}\left[-2  dx^+ \left(dx^--\frac12 \Omega_{ij}x^idx^j\right)  +  dx^i  dx^i \right] \, .\end{equation} 
In fact this is simply  six-dimensional Minkowski space,  as one would expect to find in the traditional flow to the boundary of $AdS_7$ (in general the boundary is  only defined up to a conformal class). In this case   the coordinate $x^+$ has a finite range  $x^+\in (- \pi R_+,\pi R_+)$.  However  the metrics
  $\dd s^2_{\phi\to\infty}$ and $\dd s^2_{\rho\to\infty}$ only differ by a conformal factor and the M5-brane theory, being conformal, is insensitive to this rescaling.\footnote{In principle, there could be a Weyl anomaly \cite{Henningson:1998gx}, but the anomaly vanishes for the metric in \eqref{metricboundary} and \eqref{minkowski}.
}  So the boundary worldvolume theories will agree.

Thus for embeddings at finite constant values of $\phi$ and $\rho$ we   obtain   two different descriptions of the M5-brane (although we don't know how to construct an action for the case of finite $\rho$), both coupled  to bulk eleven-dimensional M-theory modes. However when we take the limits   $\phi,\rho\to\infty$ the bulk modes decouple and either embedding leads to a description of the $(2,0)$ theory of M5-branes  on the six-dimensional spacetime with metric
\begin{equation}
\dd s^2  \ = \  -2  dx^+ \left(dx^--\frac12 \Omega_{ij}x^idx^j\right)  +  dx^i  dx^i  \, ,\label{gdef}
\end{equation}
where we have again performed a conformal rescaling to simplify the metric.  

From the point of view of Minksowski space the coordinate $x^+$ in (\ref{gdef}) takes values in the range $x^+\in (- \pi R_+,\pi R_+)$.
We can then  extend the range to $x^+\in [- \pi R_+,\pi R_+]$ to obtain a spacetime we refer to as  null conformally compactified Minkowski space. Note however that in this null conformally compactified spacetime $x^+$ is not necessarily periodic. However, as we have seen the original $AdS_7$ boundary corresponds to a periodic  $x^+$ with period $4\pi R_+$. We propose that we can  extend the range of $x^+$ to $[-2\pi R_+, 2\pi R_+]$ by imposing reflecting boundary conditions at $x^+=\pm \pi R_+$, making the dynamics of M5-branes on a doubled copy of Minkowski space periodic with period $4\pi R_+$, in agreement with the $AdS_7$ interpretation. 

The analogy with the ABJM construction suggests that one should consider orbifolding the fibre by modifying the periodicity of $x^+$ in \eqref{ads7} to $4 \pi R_+/k$ where $k$ is a non-negative integer. Taking $k \rightarrow \infty$ then corresponds to shrinking the fibre while setting $k=1$ recovers $AdS_7$. However, in this case the act of orbifolding does not seem to have a profound effect as in the case of M2-branes. We will comment further on this in appendix \ref{adsreduction}. 

\section{Theory at Finite Radius} \label{finite}

Let us first consider the metric at finite $\phi$ in \eqref{metricconstantphi} and reduce the abelian M5-brane theory along $x^+$. In \cite{Linander:2011jy} the action of multiple M5-branes reduced on a spacelike fibration was constructed and we will follow their methodology: namely we first consider the dimensional reduction of the abelian theory and then use supersymmetry to construct a suitable non-abelian five-dimensional action. 

\subsection{M5-brane Reduction} \label{m5finitephi}

Recall that the field content of the abelian theory is a stress tensor multiplet consisting of five scalars, a 2-form gauge field with self-dual field strength $H$ and fermions which contribute eight on-shell degrees of freedom. The equations of motion for the 2-form in the abelian theory are simply
\begin{equation}
H \ = \ \star H \, , \qquad \dd H \ = \ 0 \, .
\end{equation}
To reduce along the $x^{+}$ direction, we drop $\partial_{+}$ and
re-label in terms of five-dimensional fields as follows: 
\begin{equation}
H_{+ij} \ \equiv \ F_{ij} \, , \qquad H_{-ij} \ \equiv \ G_{ij} \, , \qquad H_{+-j} \ \equiv \ F_{-j} \, .
\end{equation}
The $\dd H=0$ equation then implies the following equations in five-dimensions:
\begin{equation}
\partial_{[i}F_{jk]} \ = \ \partial_{[-}F_{ij]} \ = \ \partial_{[i}H_{jkl]} \ = \ \partial_{-}H_{jkl}-3\partial_{[j}G_{kl]} \ = \ 0\label{bianchi} \, ,
\end{equation}
and we may consequently introduce gauge potentials $A_-,A_i$ to solve the first two of these equations.
To reduce the self-duality constraint, we will first convert to tangent
space indices: 
\begin{equation}
H_{\mu\nu\rho} \ = \ e_{\mu}{}^{\hat{\mu}}e_{\nu}{}^{\hat{\nu}}e_{\rho}{}^{\hat{\rho}}H_{\hat{\mu}\hat{\nu}\hat{\rho}}\label{tantran} \, ,
\end{equation}
where $e_{\mu}{}^{\hat{\mu}}$ is a vielbein for the metric in \eqref{metricboundary}. We then obtain the following useful relations:
\begin{align}
H_{-\mu\nu} \ &= \ H_{\hat{-}\hat{\mu}\hat{\nu}} \, , \nonumber\\
H_{ijk} \ &= \ H_{\hat{i}\hat{j}\hat{k}}+\frac{3}{2}\Omega_{[i|m|}x^{m}G_{jk]} \, , \nonumber\\
F_{ij} \ &= \ H_{\hat{+}\hat{i}\hat{j}}+\Omega_{[i|m}x^{m}F_{-|j]}+\frac{1}{2}e^{-\phi}G_{ij} \, .
\label{tangentfphi}
\end{align}
In the tangent space, the self-duality constraint is simply
\begin{equation}
H_{\hat{\mu}\hat{\nu}\hat{\rho}} \ = \ \frac{1}{6}\epsilon_{\hat{\mu}\hat{\nu}\hat{\rho}\hat{\omega}\hat{\sigma}\hat{\lambda}}H^{\hat{\omega}\hat{\sigma}\hat{\lambda}} \, .\label{eq:selfdual}
\end{equation}
Combining this constraint with the relations in \eqref{tangentfphi} then implies 
\begin{align}
G \ &= \ \star G \, , \nonumber\\
F_{-l} \ &= \ -\frac{1}{6}\epsilon_{ijkl}H_{ijk}+\frac{1}{2}\Omega_{im}x^{m}G_{il} \, , \nonumber\\
G \ &= \ e^{\phi}(\mathcal{F}+\star \mathcal{F}) \, ,
\label{selfddef}
\end{align}
where we have defined
\begin{equation}
\mathcal{F}_{ij} \ = \ F_{ij}-\Omega_{[i|m}x^{m}F_{-|j]} \, .
\end{equation}
We now propose an action for the gauge fields at finite $\phi$: 
\begin{equation}
\mathcal{L}_A \ = \ \frac{1}{2}F_{i-}F_{i-}+\frac{1}{4}e^{\phi}\left(\mathcal{F}_{ij}^{2}+\epsilon_{ijkl}\Omega_{mi}x^{m}F_{jk}F_{-l}\right) \, .
\end{equation}
In particular the equations of motion for $A_{-}$ and $A_i$ are 
\begin{align}
\partial_{i}F_{-i} \ &= \ -\frac{1}{2}e^{\phi}\left(\Omega_{im}x^{m}\partial_{j}\mathcal{F}_{ji}+ R_+^{-2} x^{j}F_{-j}\right) \, , \nonumber\\
\partial_{-}F_{-i} \ &= \ -e^{\phi}\left(\nabla_{j}\mathcal{F}_{ji}+\Omega_{ij}F_{-j}\right) \, ,
\end{align}
where  $\nabla_{i}=\partial_{i}-\frac{1}{2}\Omega_{ij}x^{j}\partial_{-}$. It is not difficult to show that these equations indeed follow from \eqref{bianchi} and \eqref{selfddef}.

Let us now consider the matter fields. For the scalar fields, we have
\begin{equation}
\mathcal{L}_{X} \ = \ -\frac{1}{2}\partial_{\mu}X^{I}\partial^{\mu}X^{I} \ = \ -\frac{1}{2}\eta^{\hat{\mu}\hat{\nu}}\left(e^{-1}\right)_{\hat{\mu}}{}^{\mu}\partial_{\mu}X^{I}\left(e^{-1}\right)_{\hat{\nu}}{}^{\nu}\partial_{\nu}X^{I} \, .
\label{scalareq}
\end{equation}
Plugging \eqref{metricconstantphi} into \eqref{scalareq}, dropping $x^+$ derivatives, and rescaling fields to get canonical kinetic terms then gives the following scalar lagrangian:
\begin{equation}
\mathcal{L}_{X} \ = \ -\frac{1}{2}\nabla_{i}X^{I}\nabla_{i}X^{I}-\frac{1}{2}e^{-\phi}\partial_{-}X^{I}\partial_{-}X^{I} \, .
\end{equation}
Similarly, for the fermions we have
\begin{equation}
\mathcal{L}_{\Psi} \ = \ \frac{\ii}{2}\bar{\Psi}\Gamma^{\mu}\left(\partial_{\mu}+\frac{1}{4}\omega_{\mu}^{\hat{\nu}\hat{\rho}}\Gamma_{\hat{\nu}\hat{\rho}}\right)\Psi \ = \ \frac{\ii}{2}\bar{\Psi}\Gamma^{\hat{\mu}}\left(e^{-1}\right)_{\hat{\mu}}{}^{\mu}\left(\partial_{\mu}+\frac{1}{4}\omega_{\mu}^{\hat{\nu}\hat{\rho}}\Gamma_{\hat{\nu}\hat{\rho}}\right)\Psi \, ,
\label{fermieq}
\end{equation}
where $\Gamma_{012345}\Psi=-\Psi$, and $\bar{\Psi}=\Psi^T C$. After plugging in \eqref{metricconstantphi} and dropping $\partial_+$ terms we obtain\footnote{From this point forward, the derivatives acting on fermions do not include spin connection terms. Also note that the fermionic mass-like term $\sim \bar\Psi \Gamma_- ( \Omega \cdot \Gamma ) \Psi$ vanishes via duality arguments.}
\begin{equation}
\mathcal{L}_{\Psi} \ = \ -\frac{\ii}{2}\bar{\Psi}\Gamma_{+}\partial_{-}\Psi+\frac{\ii}{2}\bar{\Psi}\Gamma_{i}\nabla_{i}\Psi+\frac{\ii}{4}e^{-\phi}\bar{\Psi}\Gamma_{-}\partial_{-}\Psi \, .\end{equation}
 
\subsection{Non-abelian Theory}

The lagrangian $\mathcal{L}_{A}+\mathcal{L}_{X}+\mathcal{L}_{\Psi}$ we obtained in the previous subsection can be non-abelianized as follows (all geometric quantities are those of flat space):
\begin{align}
S_\phi \ = \ \frac{1}{g^2_{YM}}{\rm tr}\int d^4x \, dx^- \bigg\{& 
 -\frac{1}{2}\nabla_{i}X^{I}\nabla_{i}X^{I}-\frac{1}{2}e^{-\phi} D_{-}X^{I} D_{-}X^{I}-\frac{1}{4} e^{-\phi}\left[X^{I},X^{J}\right]^{2} \nonumber\\
&+\frac{1}{2}F_{i-}F_{i-}+\frac{1}{4}e^{\phi}\left(\mathcal{F}_{ij}^{2}+\epsilon_{ijkl}\Omega_{mi}x^{m}F_{jk}F_{-l}\right)\nonumber\\
&-\frac{\ii}{2}\bar{\Psi}\Gamma_{+}D_{-}\Psi+\frac{\ii}{2}\bar{\Psi}\Gamma_{i}\nabla_{i}\Psi+\frac{\ii}{4}e^{-\phi}\bar{\Psi}\Gamma_{-}D_{-}\Psi\nonumber\\
&-\frac{1}{2}\bar{\Psi}\Gamma_{+}\Gamma^{I}\left[X^{I},\Psi\right]-\frac{1}{4} e^{-\phi}\bar{\Psi}\Gamma_{-}\Gamma^{I}\left[X^{I},\Psi\right]\bigg\} \, , 
\label{constantphiaction}
\end{align}
where $g^2_{YM}$ is a coupling constant with dimensions of length,
\begin{equation}
\nabla_i \ = \ D_i - \frac 12 \Omega_{ij}x^jD_- \, ,
\end{equation}
and $D_{\mu}X=\partial_{\mu}X-\ii [A_{\mu},X]$. 
To describe $N$ M5-branes we  take the gauge group to be $U(N)$.

This model enjoys the following supersymmetries:
\begin{align}
\delta X^I \ =& \ \ii\bar\epsilon^{(0)}_- \Gamma^I\Psi \, ,\nonumber\\
\delta A_i \ =& \ \ii\bar\epsilon^{(0)}_-\Gamma_i\Gamma_+\Psi + \frac{\ii}{2}\Omega_{ij}x^j\bar\epsilon^{(0)}_-\Gamma_{-+}\Psi \, ,\nonumber\\
\delta A_- \ =& \ \ii\bar\epsilon^{(0)}_-\Gamma_{-+}\Psi \, , \nonumber\\
\delta \Psi \ =& \ -\Gamma_+\Gamma^ID_-X^I\epsilon^{(0)}_- + \Gamma_i\Gamma^I\nabla_i X^I\epsilon^{(0)}_-\nonumber\\
&\ - \Gamma_{i}\Gamma_{+-}F_{-i}\epsilon^{(0)}_- -\frac{\ii}{2}\Gamma_{+}\Gamma^{IJ}[X^{I},X^{J}]\epsilon^{(0)}_- -\frac{e^{\phi}}{2}\Gamma_{ij}\Gamma_+\mathcal{F}_{ij}\epsilon^{(0)}_- \, ,
\end{align}
where $\epsilon_-^{(0)}$ is constant spinor satisfying $\Gamma_{012345}\epsilon =\epsilon$ and $\Gamma_{05}\epsilon^{(0)}_- = -\epsilon^{(0)}_-$ so that the theory has eight supercharges. The spinor $\epsilon = \epsilon^{(0)}_-$ is a solution to the Killing spinor equation for the metric in \eqref{metricconstantphi}, as described in appendix \ref{kseapp}. Note that there are also non-constant solutions to the Killing spinor equations  but we do not expect them to correspond to symmetries, as the theory is not conformal. This is also suggested by the holographic argument in section \ref{holographic}.

Further note also that the metric in \eqref{metricconstantphi} is invariant under the following rescaling of the coordinates, including $\phi$:
\begin{equation}
x^+ \ \rightarrow \ x^+ \, , \qquad x^{-} \ \rightarrow \ \lambda^{2}x^{-} \, , \qquad x^{i}\ \rightarrow \ \lambda x^{i} \, ,  \qquad  e^{\phi} \ \rightarrow \ \lambda^{-2}e^{\phi} \,  ,
\label{coordscaling}
\end{equation}
where $\lambda$ is a constant.
As a consequence, the action (\ref{constantphiaction}) is invariant under the scalings
\begin{align}
X^{I} \ \rightarrow \ \lambda^{-2}X^{I}\, , \qquad A_{-} \ \rightarrow \ \lambda^{-2}A_{-}\, , \qquad A_{i} \ \rightarrow \ &\lambda^{-1}A_{i} \, , \nonumber \\ 
	\Psi_+ \ \rightarrow \ \lambda^{-3}\Psi_+ \, , \qquad {\Psi}_- \ \rightarrow \ \lambda^{-2} {\Psi}_- \, , \label{fieldscaling}
\end{align}
but only if we also transform  the radial coordinate according to
\begin{equation}
e^{\phi} \ \rightarrow \ \lambda^{-2}e^{\phi} \, .
\label{phiscaling}
\end{equation}
Moreover, invariance of the supersymmetry variations implies that
\begin{equation}
\epsilon_+ \ \rightarrow \ \epsilon_+\, , \qquad \epsilon_- \ \rightarrow \ \lambda {\epsilon}_- \, .
\label{susyscaling}
\end{equation}
Since \eqref{phiscaling} corresponds to moving the branes along the $\phi$ direction, the scaling symmetry is broken if we hold $\phi$ fixed. 

Remarkably, the theory we obtain from dimensionally reducing the M5-brane theory at finite $\phi$ corresponds to an $\Omega$-deformation of the model considered in \cite{Lambert:2019evt} which was obtained by non-Lorentzian rescaling five-dimensional super-Yang-Mills. While for $\Omega_{ij}=0$ and $\phi=0$ we recover five-dimensional maximally supersymmetric  euclidean Yang-Mills as studied in \cite{Hull:2014cxa}.  In the present context, the non-Lorentzian rescaling is controlled by the radial coordinate $\phi$. In the next section, we will show that in the limit $\phi \rightarrow \infty$ the supersymmetry becomes enhanced and the theory becomes invariant under the rescalings in \eqref{coordscaling} and \eqref{fieldscaling}. In this sense, the boundary theory can be thought of as the fixed point of a classical RG flow induced by taking $\phi \rightarrow \infty$.

Let us make some further comments. First we need to determine the coupling $g^2_{YM}$. By comparing to the case $\Omega_{ij}=0$ we can impose the usual relation for the gauge coupling of five-dimensional super-Yang-Mills in terms of the period $x^+\sim x^++4\pi R_+$ of the compactification:
\begin{equation}
g^2_{YM} \ = \ 8\pi^2 R_+\ ,
\end{equation}
which is valid for spacelike or timelike reductions ({\it e.g.}\ see \cite{Hull:2014cxa}). 

Lastly we note that the action has a Chern-Simons-like term
\begin{equation}
-\frac{e^\phi}{4}\epsilon_{ijkl}\Omega_{mi}x^{m}{\rm tr}(F_{jk}F_{-l} )
=\frac{e^\phi}{4} \Omega_{mi}x^mdx^i \wedge {\rm tr}(F\wedge F) 
\cong \frac{e^\phi}{2}  \Omega  \wedge {\rm tr}\left(A\wedge dA + \frac{2i}{3}A\wedge A\wedge A\right)\ ,
\end{equation}
where $\Omega = \tfrac12\Omega_{ij}dx^i\wedge dx^j$ and $A$ is viewed as a 1-form in $1+4$ dimensions with components $(A_-,A_i)$.  Normally such a term is associated with a quantization of the coupling constant. However that does not seem to be the case here because the form of the action on the left hand side is manifestly gauge invariant, without compromising other symmetries. In the ABJM theory, quantization of the Chern-Simons coupling corresponded to performing an orbifold on the gravity side. Although this does not appear to be required in our context, we consider performing such an orbifold in appendix \ref{adsreduction}.


\section{Boundary Theory} \label{boundary}

In this section, we will take the $\phi \rightarrow \infty$ limit of the theory derived in the previous section and show that the supersymmetry becomes enhanced to 24 supercharges, providing a holographic realization of the classical RG flow mechanism proposed in \cite{Lambert:2019evt}. We will also show that the resulting theory can be obtained from a null reduction of the M5-brane theory at the boundary.

\subsection{M5-brane Reduction}

We will first reduce the abelian M5-brane theory along $x^+$ in \eqref{metricboundary}. The analysis is very similar to the one in section \ref{m5finitephi}. In fact, all that changes is the last line in equations \eqref{tangentfphi} and \eqref{selfddef}:
\begin{align}
F_{ij} \ &= \ H_{\hat{+}\hat{i}\hat{j}}+\Omega_{[i|m}x^{m}F_{-|j]} \, , \nonumber\\
\mathcal{F} \ &= \ -\star \mathcal{F} \, .
\label{selfd}
\end{align}
Let us now propose a lagrangian which encodes these
dimensionally reduced equations of motion for the gauge fields:
\begin{equation}
\mathcal{L}_A \ = \ \frac{1}{2}F_{-i}F_{-i}+\frac{1}{2}G_{ij}\mathcal{F}_{ij} \, ,
\end{equation}
where  $G_{ij}$ is self-dual (but not subject to a Bianchi identity). The anti-self-duality of $\mathcal{F}$ immediately
follows from varying the action with respect to $G_{ij}$, which is
essentially a Lagrange multiplier. From the point of view of the M5-brane, $G_{ij}$ can be identified with $H_{ij-}$. 
Furthermore, the equations of motion
for $A_{-}$ and $A_i$ are 
\begin{align}
\partial_{i}F_{-i}+\frac{1}{2}\partial_{i}\left(G_{ij}\Omega_{jk}x^{k}\right) \ = \ 0 \, , \nonumber\\
\partial_{-}\left(F_{-i}+\frac{1}{2}G_{ij}\Omega_{jk}x^{k}\right)-\partial_{j}G_{ij} \ = \ 0 \, ,
\end{align}
which indeed follow from \eqref{bianchi} and the second line of \eqref{selfddef}. It remains to derive a lagrangian for matter fields, but this is also straightforward. In particular, plugging \eqref{metricboundary} into \eqref{scalareq} and \eqref{fermieq}, and dropping $\partial_+$ terms gives
\begin{equation}
\mathcal{L}_{X} \ =\ -\frac{1}{2}\nabla_{i}X^{I}\nabla_{i}X^{I} \, ,
\end{equation}
and
\begin{equation}
\mathcal{L}_{\Psi} \ = \ -\frac{\ii}{2}\bar{\Psi}\Gamma_{+}\partial_{-}\Psi+\frac{\ii}{2}\bar{\Psi}\Gamma_{i}\nabla_{i}\Psi \, .
\label{boundferm}
\end{equation}

Our next step would be to find a non-abelian extension of the lagrangian 
\begin{equation}\label{abelianL}
{\cal L} = {\cal L}_A+{\cal L}_X +{\cal L}_\Psi\ ,
\end{equation}
 and supersymmetrize it. However, rather than do this we will instead consider obtaining the theory as a limit of the finite $\phi$ action along the lines presented in \cite{Lambert:2019evt}. The final result will be the same.

\subsection{Classical RG Flow}

Let us now carefully take $\phi \rightarrow \infty$ of the theory constructed in section \ref{finite}. Taking $\phi\to\infty$ in \eqref{constantphiaction} will remove the terms which scale like $e^{-\phi}$, but the following terms diverge in this limit:  
\begin{equation}
\mathcal{L}_{\mathcal{F}^{2}} \ = \ \mathcal{F}_{ij}^{2}+\epsilon_{ijkl}\Omega_{mi}x^{m}F_{jk}F_{-l} \, .\label{divergent}
\end{equation}
In fact we can write this as
\begin{align}
\mathcal{L}_{\mathcal{F}^{2}} \ = \ 2{\cal F}_{ij}^{+}{\cal F}_{ij}^{+}-\frac{1}{2}\varepsilon_{ijkl}F_{ij}F_{kl}  \, ,
\end{align}
where ${\cal F}^{+}_{ij}$ is the self-dual part of ${\cal F}_{ij}$.
Note that the second term is a total derivative. Thus we see that finiteness of the dynamics in the limit $\phi\to\infty$ imposes the constraint
\begin{equation}
{\cal F}^{+}_{ij} \ = \ 0\, ,
\end{equation}
{\it i.e.}\ ${\cal F}$ is anti-self-dual. Hence, the divergent terms can be cancelled by introducing a self-dual Lagrange multiplier $G_{ij}$  which imposes that $\mathcal{F}$ is anti-self-dual. 

We then arrive at the following action:
\begin{align}
\label{boundaction}
S \ = \ \frac{1}{g^2_{YM}}{\rm tr}\int d^4x \, dx^-   \bigg\{&  \frac12 F_{-i}F_{-i}+\frac12 \mathcal{F}_{ij}G_{ij}-\frac12 \nabla_i X^I\nabla_i X^I \nonumber \\
& -\frac{\ii}{2}\bar\Psi\Gamma_+D_-\Psi + \frac{\ii}{2}\bar\Psi\Gamma_i\nabla_i\Psi -\frac12\bar\Psi\Gamma_+\Gamma^I [X^I,\Psi] \bigg \} \, ,
\end{align}
with $G_{ij} = \frac12 \varepsilon_{ijkl}G_{kl}$. This is therefore the natural choice for the non-abelian generalization of (\ref{abelianL}).  For $U(N)$ gauge group it  describes $N$ M5-branes on  the six-dimensional space with the metric \eqref{gdef} and the restriction that $x^+$ is periodic with period $4\pi R_+$. This in turn corresponds to two copies of a null compactified Minkowski spacetime with $x^+\in[0,2\pi R_+]$ along with reflecting boundary conditions at $x^+=2\pi R_+$. Let us emphasize that this theory was obtained by a classical RG flow, along which the Yang-Mills coupling remains constant.

This is an $\Omega$-deformation of the non-Lorentzian theory constructed in \cite{Lambert:2018lgt} and obtained as a similar RG flow in \cite{Lambert:2019evt}, which can be identified as a null reduction of the M5-brane on ${\mathbb R}^{1,5}$. One can check that it is invariant under the following supersymmetries:
\begin{align}
\delta X^I \ =& \ \ii \bar\epsilon \Gamma^I\Psi \, , \nonumber\\
\delta A_i \ =& \ \ii \bar\epsilon\Gamma_i\Gamma_+\Psi + \frac{\ii}{2}\Omega_{ij}x^j\bar\epsilon\Gamma_{-+}\Psi \, , \nonumber\\
\delta A_- \ =& \ \ii \bar\epsilon\Gamma_{-+}\Psi \, , \nonumber\\
\delta G_{ij} \ =& \ \frac{\ii}{2}\bar\epsilon\Gamma_k\Gamma_{ij}\Gamma_-\nabla_k\Psi -\frac{\ii}{2}\bar\epsilon\Gamma_+\Gamma_-\Gamma_{ij}D_-\Psi+ \frac12 \bar\epsilon \Gamma_+\Gamma_-\Gamma_{ij}\Gamma^{I}[X^I,\Psi] - 3 \ii \bar\eta \Gamma_- \Gamma_{ij} \Psi \, , \nonumber\\
\delta \Psi \ =& \ -\frac14  \Gamma_{ij}\Gamma_-\mathcal{F}_{ij}\epsilon -\Gamma_+\Gamma^ID_-X^I\epsilon + \Gamma_i\Gamma^I\nabla_i X^I\epsilon - \Gamma_{i}\Gamma_{+-}F_{-i}\epsilon \, ,\nonumber\\
 &\  -\frac14\Gamma_{ij}\Gamma_+G_{ij}\epsilon -\frac{\ii}{2}\Gamma_{+}\Gamma^{IJ}[X^{I},X^{J}]\epsilon - 4 X^I \Gamma^I \eta \, ,
\label{susy1}
\end{align}
where $\epsilon$ and $\eta$ parametrise the $x^+$-independent solutions to the Killing spinor equation as described in appendix \ref{kseapp} ({\it i.e.}\ types II,III,IV). Note that there are 24 independent solutions, all of which satisfy  $\Gamma_{012345}\epsilon =\epsilon$. The constant solutions satisfy the further constraint $\Gamma_{05}\epsilon = -\epsilon$ but in addition we  find 16  superconformal symmetries where $\epsilon$ is not constant.
 
Finally it is easy to see that the action in \eqref{boundaction} is invariant under the following Lifshitz rescaling of the fields:
\begin{align}
&X^{I} \ \rightarrow \ \lambda^{-2}X^{I}\, , \qquad A_{-} \ \rightarrow \ \lambda^{-2}A_{-}\, , \qquad A_{i} \ \rightarrow \ \lambda^{-1}A_{i} \, , \nonumber \\ 
	&\Psi_+ \ \rightarrow \ \lambda^{-3}\Psi_+ \, , \qquad {\Psi}_- \ \rightarrow \ \lambda^{-2} {\Psi}_- \, , \qquad G_{ij} \ \rightarrow \ \lambda^{-4}G_{ij} \, . \label{fieldscaling2}
\end{align} 
In this sense, the theory has conformal symmetry even though it has a dimensionful coupling. 

\section{Conclusion} \label{conclusion}

In this paper we constructed the non-abelian field theories corresponding to $N$ M5-branes compactified on a timelike fibration of $AdS_7$ over $\tilde{\CP}^3$ which becomes null as we approach the boundary. In particular, the timelike reduction was constructed by considering brane embeddings at finite $\tilde{\CP}^3$ radius and turns out to be a non-abelian field theory with eight supersymmetries which resembles an $\Omega$-deformation of five-dimensional euclidean super-Yang-Mills. We then took the radius to infinity inducing a classical RG flow whose fixed point has an additional 16 superconformal symmetries and a Lifshitz scaling symmetry. Moreover, we showed that the fixed point theory describes the $(2,0)$ theory reduced along a null direction in a conformal compactification of Minkowski space corresponding to the boundary of $AdS_7$.

A number of questions immediately present themselves. The first is to identify the precise symmetry group of the five-dimensional fixed point theory and understand how it is related to symmetries of the bulk theory (see \cite{Taylor:2015glc} for a review of Lifshitz holography). We expect the bosonic subgroup to be $SU(3,1) \times SO(5)$, since these are the bulk isometries after reducing along the fibre. Once the superconformal symmetry is identified, it would interesting to check if the theory is in fact UV finite. A closely related question is to understand the moduli space dynamics of the fixed point theory. The equations of motion for null reductions of the M5-brane theory without an $\Omega$ deformation were previously shown to reduce to motion on instanton moduli space \cite{Lambert:2011gb}, familiar from the DLCQ description of the $(2,0)$ theory \cite{Aharony:1997th,Aharony:1997an}. In this description, the six-dimensional theory is recovered by taking the number of instantons to infinity but the moduli space is singular because of small instantons, which leads to conceptual and technical difficulties. It would be interesting to repeat this analysis for the model presented in this paper and see if the $\Omega$-deformation helps regulate these singularities.  

Lastly our results suggest a new paradigm for holography where the emergence of time appears to be a non-perturbative phenomenon, as studied in \cite{Hull:1998vg,Hull:2014cxa}. Ultimately, it would interesting to see if this is a generic consequence of M-theory or just a peculiar feature of our construction.

\section*{Acknowledgements}

N.~Lambert and P.~Richmond were supported   by STFC grant ST/L000326/1 and A.~Lipstein by the Royal Society as a Royal Society University Research Fellowship holder. The authors would like to thank Seok Kim, Sunil Mukhi and Alessandro Tomasiello for sharing their unpublished notes.

\appendix

\section{Conventions} \label{conventions}

\subsection{Indices}
The following table summarizes the different sets of indices we use throughout the body of this paper.
\begin{center}
\begin{tabular}{|l|l|l|}
	\hline &&\\
	$M,N,P,\ldots$ & $0,\ldots,10$ & eleven-dimensional frame indices \\&&\\
	$s,t,u\ldots$ & $  0,\ldots, {6}$ & seven-dimensional coordinate indices \\&&\\
	$\hat{s},\hat{t},\hat{u},\ldots$& $0,\ldots,6$ &  seven-dimensional frame indices \\&&\\	
	$\mu,\nu,\rho,\ldots$ & $  0,\ldots, {5}$ & six-dimensional worldvolume coordinate indices \\&&\\
	$\hat \mu,\hat \nu,\hat \rho,\ldots$& $0,\ldots,5$ &  six-dimensional  worldvolume frame indices \\&&\\
	$I,J,K,\ldots$ & $6,\ldots,10$ & transverse Spin$(5)$ R-symmetry indices \\&&\\
	$i,j,k,\ldots$ & $1,\ldots,4$ & worldvolume coordinate and frame indices \\&&\\
	$+,-$ & & worldvolume lightcone frame indices \\&&\\
	\hline
\end{tabular}
\end{center}

\subsection{Gamma matrix conventions}
The coordinate independent Clifford matrices of Spin$(1,10)$ are denoted $\Gamma_M$ and satisfy
\begin{align}
	\{ \Gamma_M , \Gamma_N \} \ = \ 2 \eta_{MN} \, , \qquad \Gamma_M^{\mathrm{T}} \ = \ - C \Gamma_M C^{-1} \, .
\end{align}
In addition, the charge conjugation matrix satisfies $C^{\mathrm{T}} = - C , C^\dagger = - C , C^{-1} = - C$ and we may take it to be $C=\Gamma_0$. The antisymmetric product of Clifford matrices is defined with weight one:
\begin{align}
	\Gamma_{M_1 \ldots M_n} \ = \ \Gamma_{[M_1} \Gamma_{M_2} \cdots \Gamma_{M_n]} \, .
\end{align}

We now decompose Spin$(1,10)$ into Spin$(1,5)\times$Spin$(5)$ by splitting the eleven-dimensional index into worldvolume and transverse indices: $\Gamma_M = ( \Gamma_{\hat{\mu}} , \Gamma_I )$ such that
\begin{align}
	\{ \Gamma_{\hat{\mu}} , \Gamma_I \} \ =& \ 0 \, , \qquad \{ \Gamma_{\hat{\mu}} , \Gamma_{\hat{\nu}} \} \ = \ 2 \eta_{{\hat{\mu}}{\hat{\nu}}} \, , \qquad \{ \Gamma_I , \Gamma_J \} \ = \ 2 \delta_{IJ} \, .
\end{align}
The fermions we introduce are single 32-component Majorana spinors and satisfy
\begin{align}
	\Gamma_{012345} \Psi \ =& \ - \Psi \, , \\
	\Gamma_{012345} \epsilon \ =& \ + \epsilon \, , \\
	\Gamma_{012345} \eta \ =& \ - \eta \, .
\end{align} 
In going to the light-cone we define
\begin{align}
	\Gamma_+ \ = \ \tfrac{1}{\sqrt{2}} ( \Gamma_0 + \Gamma_5 ) \, , \qquad \Gamma_- \ = \ \tfrac{1}{\sqrt{2}} ( \Gamma_0 - \Gamma_5 ) \, .
\end{align}
Then we can rewrite the projections conditions on the fermions as
\begin{align}
	\Gamma_{+-} \Gamma_{ijkl} \Psi \ =& \ + \varepsilon_{ijkl} \Psi \, , \\
	\Gamma_{+-} \Gamma_{ijkl} \epsilon \ =& \ - \varepsilon_{ijkl} \epsilon \, , \\
	\Gamma_{+-} \Gamma_{ijkl} \eta \ =& \ + \varepsilon_{ijkl} \eta \, ,
\end{align}
with $\varepsilon_{1234}=+1$. 

We also define 
\begin{align}
	\Omega \cdot \Gamma \ \equiv \ \Omega_{ij} \Gamma^{ij} \, .
\end{align}

\section{Brane Embeddings} \label{adskse}

In this appendix, we will review the solutions to the Killing spinor equations in $AdS_7$ derived in \cite{Pope:1999xg}. First note that the seven-dimensional Dirac matrices used in that paper are given by\footnote{In this appendix we use eight-component symplectic-Majorana spinors and $8 \times 8$ Clifford matrices.}
\begin{equation}
\gamma^{+} \ = \ -\ii \sigma_{3}\otimes\sigma_{3}\otimes\sigma_{3}\, , \qquad \gamma^{-}\ = \ 1\otimes1\otimes\sigma_{2}\, , \qquad \gamma^{\phi} \ = \ 1\otimes1\otimes\sigma_{1}\nonumber \, ,
\end{equation}
\begin{equation}
\gamma^{1} \ = \ \sigma_{1}\otimes\sigma_{3}\otimes\sigma_{3}\, , \quad \gamma^{2}\ = \ 1\otimes\sigma_{1}\otimes\sigma_{3}\, , \quad \gamma^{3}\ = \ \sigma_{2}\otimes\sigma_{3}\otimes\sigma_{3}\, , \quad \gamma^{4}\ = \ 1\otimes\sigma_{2}\otimes\sigma_{3}\, ,
\end{equation}
where the labels are understood to be tangent space indices. Moreover the Killing spinor equation is given by 
\begin{equation}
\partial_{s}\epsilon^{-}+\frac{1}{4}\omega_{s}^{\hat{t}\hat{u}}\gamma_{\hat{t}\hat{u}}\epsilon^{-} \ = \ -\frac{1}{2}\gamma_{s}\epsilon^{-}\, .
\end{equation}
Using the coordinates defined in section \ref{holographic}, the eight solutions obtained in \cite{Pope:1999xg} are then given by 
\begin{equation}
\epsilon_{1}^{-} \ = \ e^{\ii x^{+}}\left(\begin{array}{c}
0\\
0\\
0\\
0\\
0\\
0\\
0\\
1
\end{array}\right)\, ,\,\,\,\epsilon_{2}^{-} \ = \ \left(\begin{array}{c}
0\\
1\\
-\frac{1}{2}e^{\phi/2}\left(x_{4}+\ii x_{2}\right)\\
\frac{1}{2}e^{\phi/2}\left(x_{4}+\ii x_{2}\right)\\
-\frac{1}{2}e^{\phi/2}\left(x_{3}+\ii x_{1}\right)\\
\frac{1}{2}e^{\phi/2}\left(x_{3}+\ii x_{1}\right)\\
0\\
0
\end{array}\right)\, ,\,\,\,\epsilon_{3}^{-}=e^{\phi/2}\left(\begin{array}{c}
0\\
0\\
0\\
0\\
-1\\
1\\
0\\
0
\end{array}\right)\, ,
\end{equation}
\begin{equation}
\epsilon_{4}^{-} \ = \ \left(\begin{array}{c}
0\\
x_{3}-\ii x_{1}\\
-\frac{1}{2}e^{\phi/2}\left(x_{3}-\ii x_{1}\right)\left(x_{4}+\ii x_{2}\right)\\
\frac{1}{2}e^{\phi/2}\left(x_{3}-\ii x_{1}\right)\left(x_{4}+\ii x_{2}\right)\\
e^{-\phi/2}+e^{\phi/2}\left[\ii x^{-}-\frac{1}{4}\left(x_{1}^{2}-x_{2}^{2}+x_{3}^{2}-x_{4}^{2}\right)\right]\\
e^{-\phi/2}-e^{\phi/2}\left[\ii x^{-}-\frac{1}{4}\left(x_{1}^{2}-x_{2}^{2}+x_{3}^{2}-x_{4}^{2}\right)\right]\\
-\left(x_{4}+\ii x_{2}\right)\\
0
\end{array}\right)\, ,
\end{equation}
and the rest are determined from these by
\begin{equation}
\epsilon^-_{5}\ = \ -B\epsilon_{3}^{-*}\, , \qquad \epsilon^-_{6} \ = \ B\epsilon_{4}^{-*}\, , \qquad \epsilon^-_{7} \ = \ B\epsilon_{2}^{-*}\, , \qquad \epsilon^-_{8} \ = \ B\epsilon_{1}^{-*}\, ,
\end{equation}
where $B$ is defined by $B\gamma^{\hat{s}}B^{-1}=\left(\gamma^{\hat{s}}\right)^{*}$ and has the explicit form  $B=-\ii\sigma_{1}\otimes\sigma_{2}\otimes\sigma_{1}$. 

Noting that $\gamma^{+}\gamma^{-}\gamma^{1}...\gamma^{4}=-\gamma^{\phi}$, standard results \cite{Gutowski:1999iu} imply that  the amount of supersymmetry preserved by a brane embedding at constant $\phi$ is related to the number of Killing spinors which satisfy the chirality constraint
\begin{equation}
\gamma^{\phi}\epsilon^{-} \ = \ -\epsilon^{-}\, .
\end{equation}
For finite $\phi$, only $\epsilon_{3}^{-}$ and $\epsilon_{5}^{-}$ satisfy the constraint, so only eight supercharges are preserved after multiplying by four to take into account the Killing spinors of $S^4$. On the other hand, in the limit $\phi\rightarrow\infty$ we see that $\epsilon_{2}^{-},\epsilon_{4}^{-},\epsilon_{6}^{-},\epsilon_{7}^{-}$ also satisfy the constraint so the supersymmetry becomes enhanced to 24 supercharges. Note that the $x^{+}$-dependent solutions
$\epsilon_{1}^{-}$ and $\epsilon_{8}^{-}$ do not satisfy the constraint for any value of $\phi$.

\section{Killing Spinors} \label{kseapp}

Let us look for solutions to the conformal Killing spinor equation
\begin{equation}
\nabla_\mu \epsilon \ = \ \Gamma_\mu\eta \label{CKSE}\, ,
\end{equation}
for the six-dimensional metric in (\ref{gdef}) above, corresponding to the boundary $\phi \rightarrow \infty$ of $AdS_7$. We refer to a Killing spinor as conformal if it has a non-zero $\eta$. Explicitly evaluating this equation we find four classes of solutions:

\begin{center}
\begin{tabular}{|l|l|l|}
	\hline &&\\
	\multirow{3}{2 cm}{\rm type\ I} & $\epsilon_+ \ = \ e^{\frac{x^+}{4} \Omega\cdot\Gamma }\epsilon_+^{(0)}$ & $\eta_+ \ = \ 0$ \\&&\\
	& $\epsilon_- \ = \ 0$ & $\eta_- \ = \ -\frac{1}{16}e^{\frac{x^+}{4} \Omega\cdot\Gamma }(\Omega\cdot \Gamma)\Gamma_-\epsilon_+^{(0)}$ \\&&\\
	\hline &&\\
	\multirow{3}{2cm}{\rm type\ II} & $\epsilon_+ \ = \ 0$ & $\eta_+ \ = \ 0$ \\&&\\
	& $\epsilon_- \ = \ \epsilon_-^{(0)}$ & $\eta_- \ = \ 0$ \\&&\\ 
	\hline &&\\
	\multirow{3}{2cm}{\rm type\ III} & $\epsilon_+ \ = \ \epsilon_+^{(0)}$ & $\eta_+ \ = \ 0$ \\&&\\
	& $\epsilon_- \ = \ \frac12 x^i \Omega_{ij}\Gamma_{j}\Gamma_-\epsilon_+^{(0)}$ & $\eta_- \ = \ \frac{1}{16}( \Omega\cdot\Gamma)\Gamma_-\epsilon_+^{(0)}$ \\&&\\
	\hline &&\\
	\multirow{3}{2cm}{\rm type\ IV} & $\epsilon_+ \ = \ -\frac12 x^i  \Gamma_{i}\Gamma_+\epsilon_-^{(0)}$ & $\eta_+ \ = \ -\frac12 \Gamma_+\epsilon_-^{(0)}$ \\&&\\
	& $\epsilon_- \ = \ -\frac14 \Omega_{ik} \Gamma_{kj} x^ix^j\epsilon_-^{(0)}+x^-\epsilon_-^{(0)}$ & $\eta_- \ = \ -\frac{1}{16}( \Omega\cdot\Gamma)x^i\Gamma_i\epsilon_-^{(0)}$ \\&&\\
	\hline
\end{tabular}
\end{center}
Here we use the notation that $\pm$ indicates the eigenvalue with respect to $\Gamma_{05}$ and the superscript $(0)$ means that the  spinor is constant. In all cases we assume that $\Gamma_{012345}\epsilon=\epsilon$ and hence $\Gamma_{012345}\eta=-\eta$. Thus each type contains eight independent spinor components. Note that only type I has $x^+$ dependence.

It is instructive to take the $\Omega_{ij}=0$ limit so that the metric describes six-dimensional Minkowski space. In this case we see that the solutions above reduce to
\begin{center}
\begin{tabular}{|l|l|l|}
	\hline &&\\
	\multirow{3}{2 cm}{\rm type\ I} & $\epsilon_+ \ = \ \epsilon_+^{(0)}$ & $\eta_+ \ = \ 0$ \\&&\\
	& $\epsilon_- \ = \ 0$ & $\eta_- \ = \ 0$ \\&&\\
	\hline &&\\
	\multirow{3}{2cm}{\rm type\ II} & $\epsilon_+ \ = \ 0$ & $\eta_+ \ = \ 0$ \\&&\\
	& $\epsilon_- \ = \ \epsilon_-^{(0)}$ & $\eta_- \ = \ 0$ \\&&\\ 
	\hline &&\\
	\multirow{3}{2cm}{\rm type\ III} & $\epsilon_+ \ = \ \epsilon_+^{(0)}$ & $\eta_+ \ = \ 0$ \\&&\\
	& $\epsilon_- \ = \ 0$ & $\eta_- \ = \ 0$ \\&&\\
	\hline &&\\
	\multirow{3}{2cm}{\rm type\ IV} & $\epsilon_+ \ = \ -\frac12 x^i  \Gamma_{i}\Gamma_+\epsilon_-^{(0)}$ & $\eta_+ \ = \ -\frac12 \Gamma_+\epsilon_-^{(0)}$ \\&&\\
	& $\epsilon_- \ = \ 0$ & $\eta_- \ = \ 0$ \\&&\\
	\hline
\end{tabular}
\end{center}
Here type I and III have become degenerate. We can invert the third equation of type IV to find $\epsilon_-^{(0)}=\Gamma_-\eta^{(0)}_+$ and see that all four types correspond to the usual flat space ansatz $\epsilon = \epsilon^{(0)} + x^\mu\Gamma_\mu\eta^{(0)}_+$ for various choices of $\epsilon^{(0)}_\pm$ and $\eta^{(0)}_+$. However we  are missing the  solution  $\epsilon = x^\mu\Gamma_\mu\eta_-^{(0)}$. In particular this solution   corresponds to
\begin{equation}
\epsilon_+ \ = \ x^+\Gamma_+\eta_{-}^{(0)}\, , \qquad \epsilon_- \ = \ x^i \Gamma_i \eta_{-}^{(0)} \, , \qquad\eta_- \ = \ \eta_-^{(0)} \qquad \eta_+ \ = \ 0\, ,
\end{equation}
which is  dependent on $x^+$ and therefore will not survive the reduction on $x^+$. The   24 transformations arising from  $\epsilon^{(0)}_\pm$ and $\eta^{(0)}_+$   were shown to be a symmetries of the $\Omega_{ij}=0$ case in \cite{Lambert:2018lgt,Lambert:2019evt}.

\section{Timelike Reduction of $AdS_{7}\times S^{4}$ and a ${\mathbb Z}_k$ Orbifold} \label{adsreduction}

Let us  discuss here the reduction of eleven-dimensional M-theory over the $x^+$ coordinate to euclidean ten-dimensional type IIA string theory. The classic Freund-Rubin $AdS_{7}\times S^{4}$ solution of eleven-dimensional supergravity is
\begin{equation}
ds_{11}^{2} \ = \  ds_{AdS_{7}}^{2}+\frac{R_+^2}{4}ds_{S^{4}}^{2} \, , \qquad F_{11}^{(4)}\ = \ \frac{6}{R_+}\epsilon_{S^{4}}\label{eq:11metric} \, ,
\end{equation}
where $\epsilon_{S^{4}}$ is the volume form of the 4-sphere, and
the radius is related to the flux through the 4-sphere by
\begin{equation}
\left(\frac{R_+}{l_{p}}\right)^{3} \ = \  \pi N \, ,\label{eq:flux}
\end{equation}
where $l_p$ is the eleven-dimensional Planck length. 
As shown in \cite{Pope:1999xg}, $AdS_{7}$ is a Hopf fibration of a non-compact
three-dimensional complex projective space, $\tilde{\CP}^{3}$. 
The $AdS_{7}$ metric is 
\begin{equation}
ds_{AdS_{7}}^{2}\ = \ R_+^2\left[-  \left(d \tau + \omega\right)^{2}+ds_{\tilde{\CP}^{3}}^{2}\right]\label{eq:hopf} ,
\end{equation}
where $\tau = x^+/2R_+$ and $d\omega=J$ is the K\"ahler form of $\tilde{\CP}^{3}$. 
Recall
that $\tilde{\CP}^{3}$ is defined by
\begin{equation}
\bar{z}_{a}z^{a} \ = \ 1 \, , \qquad  z^{a}\sim e^{\ii\theta}z^{a}\label{eq:cp3} \, ,
\end{equation}
where $a=0,1,2,3$ and
\begin{equation}
\bar{z}_{a} \ = \ \eta_{ab}\bar{z}^{b} \, , \qquad \eta \ = \ {\rm diag\left(1,-1,-1,-1\right)} \, .
\end{equation}
In terms of homogeneous coordinates,
\begin{equation}
ds_{\tilde{\CP}^{3}}^{2} \ = \ -d\bar{z}_{a}dz^{a}+\omega^2 \, , \qquad \omega \ = \ \ii \bar{z}_{a}dz^{a}\, .\label{eq:cp3metric}
\end{equation}

Let us reduce on $x^+$ to a euclidean  IIA string theory \cite{Hull:1998ym} in a background given
by
\begin{equation}
ds_{11}^{2} \ = \ e^{-2\Phi/3}ds_{10}^{2}-e^{4\Phi/3}\left(d x^++A\right)^{2}\, , \qquad F_{11}^{(4)}\ = \ e^{4\Phi/3}F_{10}^{(4)}-e^{\Phi/3}F_{10}^{(3)}\wedge d x^+ \, .
\end{equation}
The equations of motion for ten-dimensional euclidean IIA supergravity are given \cite{Hull:1998vg} by varying 
\begin{align}
S_{NS} \ =& \ + \frac{1}{2\kappa^{2}}\int d^{10}x\sqrt{-g}e^{-2\Phi}\left\{R+4\left(\partial\Phi\right)^{2}+\left|F^{(3)}\right|^{2}\right\} \, , \nonumber \\
S_{R} \ =& \ -\frac{1}{4\kappa^{2}}\int d^{10}x\sqrt{-g}\left\{-\left|F^{(2)}\right|^{2}+\left|F^{(4)}\right|^{2}\right\} \, , \nonumber \\
S_{CS} \ =& \ -\frac{1}{4\kappa^{2}}\int d^{10}x \, B^{(2)}\wedge F^{(4)}\wedge F^{(4)}\label{eq:eom} \, ,
\end{align}
where $\kappa^{2}$ is proportional to Newton's constant. Notice the non-standard
signs appearing in the measure and kinetic terms for $F^{(2)}$ and
$F^{(3)}$, which are a result of reducing eleven-dimensional supergravity along
a timelike direction. Although some of the gauge fields have kinetic
terms with a non-standard sign, this is not necessarily problematic since unitarity
is not defined in euclidean signature.
 
Recall the usual relations
\begin{equation}
l_{str}^2 \ = \ \frac{l_p^3}{R_{11} } \, , \qquad g_{str} \ = \ \frac{R_{1}}{l_{str}} \, ,
\end{equation}
where $R_{11} = 2R_+$ is the radius of the eleventh dimension.
Ignoring powers of $2$ and $\pi$ we find
\begin{equation}
l_{str}  \ \sim \ \frac{l_p}{N^{1/6}}  \, , \qquad  g_{str} \ \sim \ N^{1/2} \, ,
\end{equation}
so that the string metric is 
\begin{equation}
ds_{10}^{2} \ = \ R_{str}^{2}\left(ds_{\tilde{\CP}^{3}}^{2}+\frac{1}{4}ds_{S^{4}}^{2}\right) ,
\end{equation}
with $R_{str} = N^{2/3}l_{str}$.
Thus we can trust the low energy supergravity approximation to M-theory or euclidean type IIA string theory when $N\gg 1$. But the string theory description is never weakly coupled. 

To obtain a weakly coupled euclidean type IIA description away from the boundary we could impose a further ${\mathbb Z}_k$ orbifold on the timelike $S^1$ fibre. However unlike the ABJM case this does not seem particularly natural, in part because the dual field theory has a coupling constant $g^2_{YM}\sim R_+$ with dimensions of length and which is not subject to a quantization condition. Nevertheless let us impose by hand the additional orbifold $x^+\cong x^+ + 4\pi R_+/k$ for some $k\in\{1,2,3,...\}$. This will not change the supergravity fields but does change the length of the eleventh-dimension   to $R_{11}=2R_+/k$. As a result we now find
\begin{equation}
l_{str}  \ \sim \ \frac{k^{1/2}}{N^{1/6}} l_p \, , \qquad  g_{str} \ \sim \ \frac{N^{1/2}}{k^{3/2}}\, ,\qquad R_{str} \ = \ \frac{N^{2/3}}{k} l_{str} \, .
\end{equation}
Here the string theory description is weakly coupled when $N \ll k^3$ and the ten-dimensional supergravity approximation is valid when $N\gg k^{3/2}$:
\begin{equation}
1 \ll k^{3/2} \ll N \ll k^3\, .
\end{equation}

In terms of the dual field theory the ${\mathbb Z}_k$ orbifold means that we restrict the M5-brane dynamics to be periodic under $x^+\to x^++4\pi R_+/k$. Thus for $k\in\{1,2,3,...\}$ we are looking at a subsector of the original theory. From the point of view of the worldvolume theory this means that the coupling is shifted to
\begin{equation}
g^2_{YM} \ = \ 8\pi^2 R_+/k\, .
\end{equation}
Given the usual interpretation of the instanton-solitons with instanton number $n$ as representing momentum modes with momentum $n/2R_+$ we now see that the only momentum modes are $kn/2R_+$. Thus we are projecting out all momentum modes that are not compatible with the periodicity $x^+\to x^++4\pi R_+/k$.

\end{document}